\documentclass[sigconf, screen, nonacm]{acmart}

\AtBeginDocument{%
  \providecommand\BibTeX{{%
    \normalfont B\kern-0.5em{\scshape i\kern-0.25em b}\kern-0.8em\TeX}}}
    
\setcopyright{none}

\usepackage{amssymb}
\usepackage{sidecap}
\usepackage{booktabs}
\usepackage{amsmath} 
\usepackage{makecell}
\usepackage{amsfonts}
\usepackage{bm}

\usepackage{enumitem}
\usepackage{multirow} 
\usepackage{algorithm}
\usepackage{algpseudocode}
\usepackage{xcolor}
\usepackage{url}
\usepackage{subfloat}
\usepackage{subfig}
\usepackage{pifont}

\usepackage{tikz}

\newcommand{\name}{HW-Router}

\usepackage[subtle,tracking=normal, title=normal]{savetrees}

\definecolor{Orange}{RGB}{255,127,14}
\definecolor{Blue}{RGB}{31,119,180}
\definecolor{Red}{RGB}{214,39,40}

\begin{document}
\settopmatter{printfolios=true}
\settopmatter{printacmref=false} 
\renewcommand\footnotetextcopyrightpermission[1]{}
\pagestyle{plain}

\title{HW-Router: Hardware-Aware Routing for \\Scalable Multi-LLM Serving}

% \author{Ahasan Kabir, Jiaqi Xue, Mengxin Zheng, Qian Lou}
% \affiliation{%
%   \institution{University of Central Florida}
%   \city{Orlando}
%   \state{Florida}
% \email{{ahasan.kabir, jiaqi.xue, mengxin.zheng, qian.lou}@ucf.edu}

\author{%
Ahasan Kabir \quad Jiaqi Xue \quad Mengxin Zheng \quad Qian Lou \\
University of Central Florida \\
\texttt{\{ahasan.kabir, jiaqi.xue, mengxin.zheng, qian.lou\}@ucf.edu}
}

\renewcommand{\shortauthors}{Kabir et al.}

\begin{abstract}

Modern large language model (LLM) serving platforms deploy multiple models across different GPUs, requiring routers to direct incoming queries to appropriate LLMs. However, existing routing approaches primarily rely on static model attributes such as size or FLOPs to estimate serving costs. This static cost modeling fails to capture the dynamic behavior of real deployments, where the same model can exhibit vastly different inference latencies depending on hardware type (e.g., H100 vs.\ V100), current system load (e.g., running and waiting queue lengths), and resource contention (e.g., KV-cache usage and GPU utilization). Such hardware-agnostic routing leads to suboptimal decisions, resulting in SLO violations, queue buildup, and underutilized GPUs. To address these challenges, we present \textbf{HW-Router}, a dynamic routing framework that integrates real-time hardware signals into model selection to enable accurate latency prediction and intelligent, SLO-aware routing decisions. Our approach incorporates model-specific features (architecture, size, input length) alongside hardware metrics including queue lengths, KV-cache utilization, and recent TTFT/TPOT performance, and uses a lightweight latency predictor to estimate per-model-per-GPU serving time. Evaluations across diverse workloads show that HW-Router achieves \textbf{3.4–3.9$\times$ lower end-to-end latency}, \textbf{46–48 percentage points higher SLO attainment}, \textbf{6–8$\times$ lower GPU load skew}, and a \textbf{3.1–3.4$\times$ reduction in waiting-queue fraction} compared to state-of-the-art router baselines, CARROT and IRT, with only \bm{$\sim 200 \mu s$} of additional routing overhead and no loss in output quality. These results highlight the importance of real-time hardware feedback for scalable, predictable, and well-balanced multi-LLM serving. Code is available at \url{https://github.com/UCF-ML-Research/HW-Router}.
\end{abstract}

\maketitle

% \fi
%\IEEEpeerreviewmaketitle

\section{Introduction}
Large language models (LLMs), including GPT-5~\cite{GPT5}, Gemini~\cite{comanici2025gemini}, Claude~\cite{claude3}, Qwen~\cite{yang2025qwen3}, and DeepSeek-V3~\cite{liu2024deepseek}, achieve strong performance across a wide range of language understanding, reasoning, and generation tasks. As these models evolve, their capabilities have also become increasingly specialized; some focus on coding, while others are optimized for mathematical reasoning, creative writing, or domain-specific knowledge. This specialization has made it common for serving platforms to deploy multiple LLMs and route each user query to the most appropriate model. For example, GPT-5~\cite{GPT5} internally selects the best-suited OpenAI model to handle a given request. Currently, predominant practices utilize homogeneous GPU clusters (e.g., Orca~\cite{yu2022orca}, vLLM~\cite{kwon2023efficient}) or heterogeneous GPU clusters (e.g., Helix~\cite{mei2025helix}, KServe~\cite{KServe}) to serve the incoming requests. However, the increasingly large model sizes and high computational requirements make it challenging to serve multiple LLMs cheaply and efficiently.

\begin{figure}[t!]
    \centering
    \includegraphics[width=\linewidth]{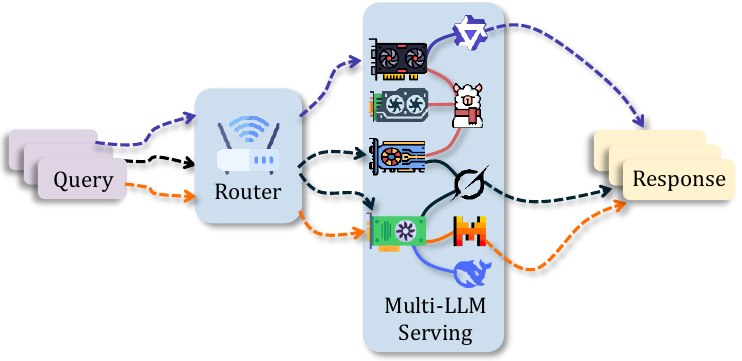}
    \vspace{-0.2in}
    \caption{LLM Router for Multi-LLM Serving. Queries are assigned to different LLMs deployed on multiple GPUs through the router. Because the LLMs run on different hardware, the router must consider each model’s capabilities as well as its expected inference cost, which depends on dynamic factors such as GPU type, queue lengths, and KV-cache utilization.}
    \label{fig:router_overview}
    \vspace{-0.2in}
\end{figure}

Recently, LLM routers have become an essential component in multi-LLM systems by automatically assigning each user query to the most appropriate model based on its capabilities and costs (see Figure~\ref{fig:router_overview}). A router estimates the expected response quality and inference cost for each model on a given query, and then selects the model that provides the best trade-off.

\begin{figure}[h!]
    \centering
    \includegraphics[width=\linewidth]{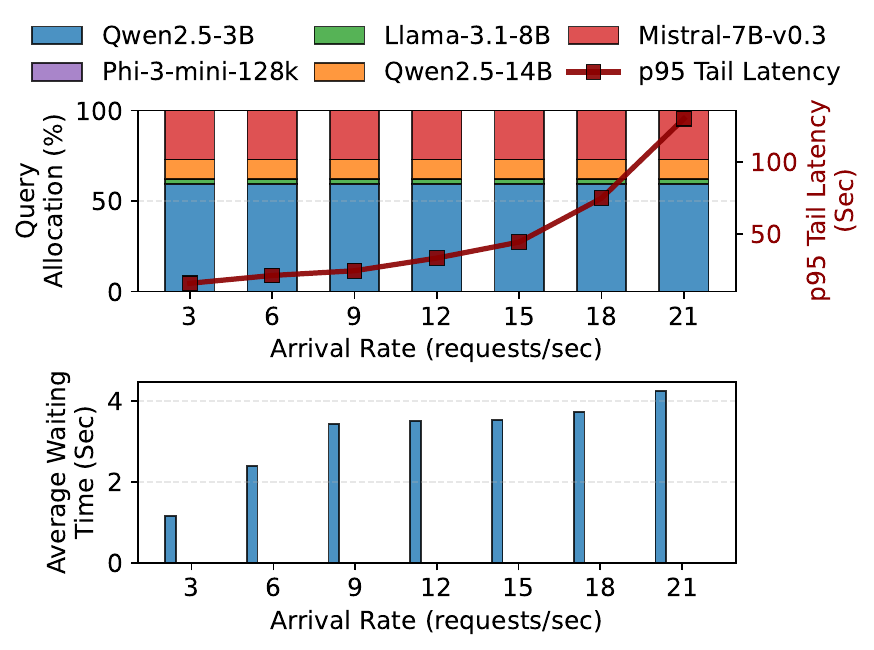}
    \vspace{-0.25in}
    \caption{Impact of static cost modeling on routing performance using existing router approach. Top: Query allocation remains unchanged. This leads to rapidly increasing p95 tail latency at higher workloads. Bottom: The average waiting time for Qwen2.5-3B grows significantly with load.}
    \label{fig:motivation}
    \vspace{-0.2in}
\end{figure}

Existing routing approaches primarily rely on static model attributes, such as parameter count or FLOPs, to approximate inference cost. For instance, RouteLLM~\cite{ongroutellm} assumes larger LLMs always have higher latency than smaller one. This static cost modeling ignores the dynamic behavior of real serving environments, where latency depends heavily on system load and resource contention. To illustrate this limitation, we evaluate the state-of-the-art CARROT~\cite{somerstep2025carrot} router in a setting where five LLMs are deployed on two GPUs (see details in Table~\ref{tab:deployment}), and vary the request arrival rate (workload, i.e., requests per second) using the same set of input queries. As shown in Figure~\ref{fig:motivation} (top), CARROT produces the exact same allocation distribution under different arrival rates, with the majority of queries consistently routed to Qwen2.5-3B. This happens because CARROT treats both quality and cost as static model attributes and bases its decisions solely on them, and Qwen2.5-3B offers the best trade-off on this set of queries.

However, a real system exhibits substantial load-dependent performance discrepancies, because GPU resources such as KV cache and memory allocated to each model are limited and dynamically changing. As shown in Figure~\ref{fig:motivation}~(bottom), the average waiting time for Qwen2.5-3B increases sharply under higher workloads, while other models experience near-zero queueing delay. This imbalance indicates poor utilization of the overall model pool, which in turn leads to a dramatic increase in latency of the whole system (Figure~\ref{fig:motivation}, top). These results highlight that static cost assumptions can cause routers to over-concentrate traffic onto a single model, ultimately degrading latency and violating SLOs.

To address these challenges, we propose HW-Router, the first hardware-aware routing framework that incorporates real-time system and hardware signals into model selection. Our insight is that an ``expensive'' model may in practice be more efficient than a ``cheaper'' one when hardware conditions are favorable. For instance, when the GPU running the ``expensive'' model is idle, has lower queue pressure, or provides higher effective throughput than the overloaded GPU serving the ``cheaper'' model. 

\begin{figure*}[ht!]
    \centering
    \includegraphics[width=\linewidth]{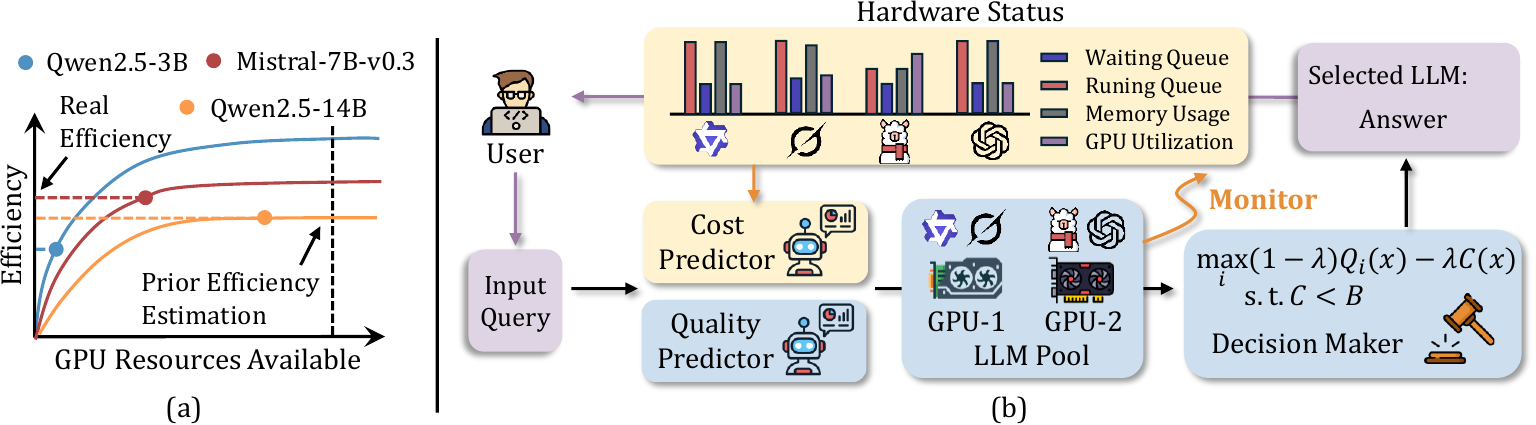}
    \caption{(a) Model efficiency as a function of available GPU resources. Prior routing methods use static cost estimates that assume constant efficiency, while actual performance varies significantly with memory availability. Larger model, e.g., Qwen2.5-14B, may have better efficiency than smaller model, e.g., Qwen2.5-3B if it has more resources. (b) HW-Router overview with four key components: Hardware Monitor tracks real-time metrics (queue lengths, memory usage, GPU utilization) across the LLM pool; Cost and Quality Predictors estimate latency and response quality for each model-GPU pair; Decision Maker selects the optimal assignment that maximizes quality-cost trade-off while satisfying SLO constraints.}
    \label{fig:overview}
\end{figure*}

As shown in Figure~\ref{fig:overview}~(a), existing routers estimate inference cost solely from model size, implicitly assuming that this relationship remains constant. However, this assumption only holds under identical system and hardware conditions, where Qwen2.5-3B appears more efficient than Mistral-7B-v0.3 and Qwen2.5-14B. In real serving environments, this efficiency ordering breaks down: when heavy traffic is routed to Qwen2.5-3B, its available memory decreases and its efficiency degrades sharply, eventually becoming slower than the much larger but idle Qwen2.5-14B.

This observation motivates HW-Router's dynamic approach. Instead of relying on static estimates, HW-Router predicts inference latency for each model by integrating model features, hardware characteristics, and real-time system metrics including queue lengths and KV-cache utilization. Our lightweight latency predictor captures the actual efficiency curves shown in Figure~\ref{fig:overview}~(a), adapting routing decisions to current serving conditions. By accounting for how model efficiency varies with memory availability, HW-Router prevents traffic concentration on seemingly "optimal" models, maintains balanced load distribution, and ensures predictable latency even under high load. This hardware-aware approach significantly improves both GPU utilization and tail latency compared to static routing strategies in multi-LLM deployments.

In the experiments, we train HW-Router’s latency-prediction model and evaluate it on a multi-LLM deployment of five models across two NVIDIA H100 GPUs under a sustained-overload arrival pattern. Across all settings, HW-Router delivers \textbf{3.4--3.9$\times$} lower end-to-end latency, \textbf{46--48} percentage-point higher SLO attainment, and \textbf{6--8$\times$} lower GPU load skew compared to both CARROT and IRT. HW-Router also cuts the waiting portion of the total queue by \textbf{3.1--3.4$\times$}, keeping GPUs in the \emph{running} state \textbf{83--84\%} of the time (vs.\ 44--47\% for CARROT and 61--66\% for IRT). These improvements remain stable from arrival rate 15 through 21, showing that real-time hardware awareness prevents queue buildup, eliminates GPU hotspots, and enables consistently balanced utilization even under heavy load.

\section{Background and Related Work}
\subsection{Multi-LLM Serving on GPU Clusters}
Modern LLM serving platforms increasingly deploy multiple models to support diverse tasks and user workloads. These systems run on GPU clusters that vary widely in hardware composition, ranging from homogeneous deployments with a single GPU type to heterogeneous clusters that combine multiple GPU generations such as V100, A100, and H100. Systems such as Orca \cite{yu2022orca} and vLLM \cite{kwon2023efficient} typically assume uniform hardware for simplicity and predictable performance, while others such as Helix \cite{mei2025helix} and KServe \cite{KServe} explicitly operate on mixed accelerators to improve cost efficiency and resource utilization.

Regardless of cluster composition, serving multiple LLMs presents significant challenges. Models differ in size, architecture, and memory requirements, which leads to different placement and execution constraints across GPUs. Larger models often require high-memory or high-throughput accelerators, while smaller models can run on more modest hardware. In both homogeneous and heterogeneous environments, inference latency is influenced not only by model characteristics but also by the underlying hardware type, current load, queueing effects, and multi-tenant contention. As a result, real-world LLM serving systems experience substantial variability in end-to-end latency that is not captured by static estimates of model FLOPs or parameter count.

\begin{table}[h!]
\centering
\small
\setlength{\tabcolsep}{0.5pt}
\caption{Comparison between \name{} and prior routers.}
\label{tab:related_work}

\begin{tabular}{lccccc}
\toprule
\makecell[l]{Methods} & \makecell[c]{Quality\\Modeling} & \makecell[c]{Cost\\Predicting} & \makecell[c]{Hardware\\Aware} & \makecell[c]{SLO\\Guarantee} & \makecell[c]{Load \\Balancing}\\
\midrule

FrugalGPT~\cite{chen2024frugalgpt} & \ding{51} & \ding{55} & \ding{55} & \ding{55} & \ding{55} \\

RouteLLM~\cite{ongroutellm} & \ding{51} & \ding{55} & \ding{55} & \ding{55} & \ding{55} \\

IRT-Router~\cite{song2025irt} & \ding{51} & \ding{55} & \ding{55} & \ding{55} & \ding{55} \\

CARROT~\cite{somerstep2025carrot} & \ding{51} & \ding{51} & \ding{55} & \ding{55} & \ding{55} \\\midrule
\name{} & \ding{51} & \ding{51} & \ding{51} & \ding{51} & \ding{51} \\
\bottomrule
\end{tabular}
\vspace{-0.1in}
\end{table}

\subsection{LLM Routing}
Routing is an important mechanism in multi-LLM serving systems. The objective is to select, for each user query, the model that provides the best trade-off between response quality and serving cost. Existing routing methods typically break this problem into two components: \textbf{quality modeling}, which predicts how well each model is expected to perform on the query, and \textbf{cost modeling}, which estimates the expected inference cost.

Most routers develop predictors that estimate the response quality of each model conditioned on the input query. Approaches such as FrugalGPT~\cite{chen2024frugalgpt}, RouteLLM~\cite{ongroutellm}, IRT-Router~\cite{song2025irt}, CARROT~\cite{somerstep2025carrot}, and R2-Router~\cite{xue2026r2} learn lightweight models to approximate the quality difference among candidate LLMs. These quality predictors are typically trained using supervised signals derived from model outputs. While effective at capturing task specialization across LLMs, these methods primarily focus on quality prediction and do not incorporate detailed or dynamic cost estimates. Among prior work, CARROT~\cite{somerstep2025carrot} is the only method that attempts to predict query-dependent cost by training a token-length predictor. This allows CARROT to approximate output cost more accurately than approaches that rely solely on static model properties. In contrast, FrugalGPT~\cite{chen2024frugalgpt}, RouteLLM~\cite{ongroutellm}, and IRT-Router~\cite{song2025irt} use the model size as a fixed cost proxy for every query. Such static cost assumptions fail to capture the significant variation in inference latency that arises from different input lengths, batch sizes, or runtime contention.

\noindent\textbf{Comparison with Related Work.} Table~\ref{tab:related_work} summarizes the key differences between HW-Router and existing routing approaches. The most critical distinction lies in the last three columns: hardware awareness, SLO guarantees, and load balancing. None of the existing routers consider the dynamic system states and hardware signals such as queue congestion or memory pressure, treating a heavily loaded GPU and an idle one as equivalent routing targets. Without visibility into actual serving conditions, routers may continuously direct queries to overloaded GPUs while identical models on idle hardware remain underutilized, and this issue becomes increasingly severe as deployments scale to more models and diverse hardware.

HW-Router addresses these realistic limitations by taking hardware awareness and load balancing into considerations. By incorporating GPU specifications, real-time queue states, and resource utilization into its latency prediction model, HW-Router can accurately estimate actual serving time rather than relying on static proxies. This enables true SLO-aware routing that adapts to the dynamic, heterogeneous nature of modern LLM serving infrastructure, ensuring predictable performance even under varying load conditions.

\section{HW-Router Design}
\subsection{Notation and preliminaries}
Let \((\mathcal{M,G})=\{(M_1, g_1), (M_2, g_2), ..., (M_n, g_n)\}\) denote the set of available model-GPU pairs, where each \(M_i\) represents an LLM and \(g_i\) denotes the specific GPU instance(s) on which it is hosted. Let \(\mathcal{X}=\{x_1, x_2, ..., x_m\}\) denote the set of incoming user queries. The goal of an LLM router is to assign each query \(x_i \in \mathcal{X}\) to the most suitable model–GPU pair \((M_j, g_j) \in (\mathcal{M,G})\), achieving high response quality at low inference cost under the current hardware conditions.

Each candidate model is characterized by two quantities: its estimated response quality \(Q\) and inference cost \(C\). For a given query \(x_i\), the router defines a trade-off score that balances quality and cost, i.e., \(S=(1-\lambda)\cdot Q-\lambda \cdot C\), where \(\lambda\) is a user-defined coefficient that controls cost sensitivity. The router then selects the model that maximizes this score.

\subsection{Methodology}
Figure~\ref{fig:overview} illustrates the workflow of HW-Router, which augments traditional LLM-routing with real-time hardware awareness. Instead of relying solely on offline model statistics, HW-Router dynamically adapts routing decisions based on current GPU conditions and runtime serving states. HW-Router consists of four main components:

\noindent\textit{1) Hardware Monitor:} Continuously polls metrics from each LLM server's status at 200-300ms intervals. For each model-GPU pair \((M_i, g_i)\), the monitor tracks the runtime signals $S_i$ shown in Table~\ref{tab:features}, which are cached locally and read by the router without introducing additional latency during request processing. These metrics capture three key aspects of system state: \texttt{running\_req\_count} and \texttt{waiting\_req\_count} indicate the current workload and queue pressure; \texttt{kv\_cache\_usage} reflects memory availability that constrains batch size and throughput; and \texttt{ttft\_avg} and \texttt{tpot\_avg} provide recent performance indicators that reveal how efficiently the model-GPU pair is currently operating under its load conditions.

\begin{table}[h]
\centering
\caption{Runtime metrics tracked by the hardware monitor}
\label{tab:features}
\begin{tabular}{ll}\toprule
\textbf{Metric} & \textbf{Description} \\
\midrule
\texttt{running\_req\_count} & Active requests being processed \\
\texttt{waiting\_req\_count} & Requests queued for execution \\
\texttt{kv\_cache\_usage} & KV-cache memory utilization (\%) \\
\texttt{ttft\_avg} & Recent avg. time-to-first-token \\
\texttt{tpot\_avg} & Recent avg. per output-token latency \\
\bottomrule
\end{tabular}
\end{table}

\noindent\textit{2) Cost Predictor:} 
Given a query $x$ and candidate model $M_i$ on GPU $g_i$, the cost predictor estimates three quantities:
\begin{equation}
\text{TTFT}_i(x),\text{TPOT}_i(x),N_{\text{out},i}(x) 
= f_{\text{cost}}(\text{feat}(x, M_i, S_i)),
\end{equation}
where $N_{\text{out},i}(x)$ denotes the predicted number of output tokens. The predictor takes as input the concatenation of (i) query features, (ii) one-hot encoded model/GPU identifiers, and (iii) the runtime metrics $S_i$ collected by the hardware monitor. These features are passed through a lightweight neural network with two fully connected layers (64-128 hidden units with ReLU activations) and three output heads that predict TTFT, TPOT, and output-token count. The resulting cost is computed as
\begin{equation}
C_i(x) = \text{TTFT}_i(x) + N_{\text{out},i}(x)\cdot \text{TPOT}_i(x).
\label{eq:cost}
\end{equation}
The entire forward pass takes of the cost predictor approximately $10\,\text{ms}$, ensuring negligible routing overhead while accurately capturing the complex interactions between model workloads, hardware contention, and query characteristics.

\noindent\textit{3) Quality Predictor:} Estimates expected response quality $Q_i(x)$ for each model $M_i$ given query $x$, following existing approaches~\cite{ongroutellm, somerstep2025carrot}. Quality predictions remain hardware-independent as response quality depends only on model capability.

\noindent\textit{4) Decision Maker:} Combines quality and cost predictions to select the optimal model-GPU assignment. For each query, it evaluates all feasible candidates and selects:
\begin{equation}
(M^*, g^*) = \arg\max_{M_i \in \mathcal{M}, g_i \in \mathcal{G}} (1-\lambda)\cdot Q_j(x) - \lambda \cdot C_i(x),
\end{equation}
where $C_i(x)$ represents latency defined in Equation~\ref{eq:cost}, and subject to SLO constraint $C_i(x) < B$.

\section{Experimental Setup}
All experiments are run on a single node equipped with two NVIDIA H100 PCIe GPUs
(80~GB HBM each), 8 CPU cores, and 64~GB system memory. All models are served
using \texttt{vLLM} with one process per GPU and Prometheus metrics enabled to
export per-GPU hardware signals and latency statistics.

We embed all prompts using the all-MiniLM-L6-v2 encoder~\cite{sentence-transformers-all-MiniLM-L6-v2}, the same model used in CARROT~\cite{somerstep2025carrot}, ensuring fully comparable quality and cost predictions. Our hardware-aware router introduces only $200\mu s$ of additional routing latency over CARROT on average, keeping overall decision-time efficiency effectively unchanged.

\subsection{Baselines}
We select CARROT~\cite{somerstep2025carrot} and IRT-Router~\cite{song2025irt} as our baselines because they are the Top-2 open-sourced LLM routers according to RouterArena~\cite{lu2025routerarena}.

\noindent\textbf{CARROT} employs a two-stage approach: first predicting both the performance and cost for each model-query pair, then selecting the model that minimizes a convex combination of these metrics. CARROT utilizes a fine-tuned RoBERTa models to predict response quality and response token length of each LLM for the input query.

\noindent\textbf{IRT-Router} adapts Item Response Theory to model the relationship between LLM capabilities and query attributes. It treats LLMs as "test-takers" with latent multidimensional abilities and queries as "test items" with difficulty and discrimination parameters. A neural interaction layer combines per-query relevance vectors with the gap between model abilities and query difficulties to predict performance.

For fair comparison, we use the default configurations reported in their respective papers, with all routers evaluated on the same test splits to ensure consistency.

\subsection{Model Pool and Placement}
We deploy five open-source, instruction-tuned LLMs from HuggingFace
(Phi-3-mini~\cite{abdin2024phi3technicalreporthighly}, Qwen2.5-3B~\cite{yang2025qwen3}, Mistral-7B~\cite{jiang2023mistral7b}, Llama-3.1-8B~\cite{foundation_sec_8b_instruct2025}, and Qwen2.5-14B~\cite{qwen2024qwen2.5}).
These models cover a range of parameter scales and latency characteristics,
providing a practical mix that reflects the heterogeneity seen in multi-model
serving setups. This variety allows us to evaluate routing behavior under
different computational demands without relying on a single model family.
The model--GPU placement is fixed for all experiments and is summarized in
Table~\ref{tab:gpu-layout}.

\begin{table}[h]
\centering
\small
\caption{GPU--model deployment used for both training and evaluation.}
\label{tab:gpu-layout}
\begin{tabular}{ll}
\toprule
\textbf{GPU} & \textbf{Models} \\
\midrule
H100--0 & Phi-3-mini~\cite{abdin2024phi3technicalreporthighly}, Qwen2.5-14B~\cite{yang2025qwen3} \\
H100--1 & Qwen2.5-3B~\cite{yang2025qwen3}, Mistral-7B~\cite{jiang2023mistral7b}, Llama-3.1-8B~\cite{grattafiori2024llama3herdmodels} \\
\bottomrule
\end{tabular}
\label{tab:deployment}
\vspace{-0.2in}
\end{table}

\subsection{Dataset}

We mix 6{,}000 prompts sampled from MixInstruct~\cite{jiang-etal-2023-llm} with 2{,}425 prompts from the English LongBench~\cite{bai2023longbench} tasks, yielding a combined set of 8{,}425 prompts (6{,}740 train / 1{,}685 eval). MixInstruct provides short instruction-style queries, while LongBench contains medium- and long-context inputs. We combine them to emulate realistic workload pressure: short prompts drive high concurrency, whereas long prompts stress KV-cache usage and latency. This mixture enables training and evaluating our hardware-aware router under diverse request lengths.

To obtain quality scores for each model's response, we follow the approach of CARROT~\cite{somerstep2025carrot} by employing an LLM-as-a-judge methodology. We use Qwen-3-Next-80B~\cite{yang2025qwen3} as our evaluator model, which assesses each response against golden reference answers and assigns a quality score from 0.0 (completely incorrect) to 1.0 (fully correct) in 0.1 increments. This is crucial for evaluating on open-ended instruction queries as well as mitigating errors associated with traditional automatic evaluation methods like exact match.

% \subsection{Evaluation Metrics}
% We evaluate routing performance using latency-based SLO criteria calibrated from the training data. The arrival rate $\rho$ refers to the request injection rate (requests per second) used to control system load in our experiments.

% The TTFT SLO is modeled as a linear function of the prompt length:
% \[
% \text{TTFT\_SLO}(p) = a + b \cdot p,
% \]
% where $a$ and $b$ are obtained from least-squares regression and scaled using a
% 1.2$\times$ slack factor. The TPOT SLO is defined as the $70^{\text{th}}$
% percentile of the per-token latency distribution and is incorporated into the
% end-to-end SLO definition.

% A request is considered to meet the end-to-end SLO if:
% \[
% \text{Latency} \le 
% \text{TTFT\_SLO}(p) + d \cdot \text{TPOT\_SLO},
% \]
% where $p$ and $d$ denote the prompt and output token counts. This formulation
% captures both prefill and decode latency under a unified criterion used to
% compute SLO attainment in our evaluation.

\subsection{Evaluation Metrics}

We evaluate routing performance using latency-based SLO criteria and load-balance metrics.

\noindent\textbf{Arrival rate.}
The arrival rate $\rho$ denotes the request injection rate (requests/second) used to control system load.

\noindent\textbf{SLO definition.}
The TTFT SLO is modeled as a linear function of prompt length $p$:
\begin{equation}
    \text{TTFT\_SLO}(p) = a + b p,
\end{equation}
where $a,b$ are obtained via least-squares regression and scaled by a 1.2$\times$ slack factor. The TPOT SLO is the empirical $70^{\text{th}}$ percentile of per-token latencies in the training set.

A request with $p$ input tokens and $d$ output tokens meets the SLO if
\begin{equation}
\text{Latency} \le \text{TTFT\_SLO}(p) + d \cdot \text{TPOT\_SLO}.
\end{equation}

\noindent\textbf{SLO attainment.}
The SLO attainment rate is the fraction of requests whose latencies
meet the above threshold.

\noindent\textbf{Queue composition.}
For each GPU, we measure the average number of \emph{running} requests
(actively being processed) and \emph{waiting} requests (queued but not
yet executing), which reveals how much of the system’s time is spent on
useful computation versus stalled backlog.

\noindent\textbf{Load Skew, Running Skew, and Waiting Skew.}
To quantify how evenly work is distributed across GPUs, we measure three skew metrics that capture different aspects of load imbalance. For a two-GPU system, the overall \emph{load skew} is defined as
\begin{equation}
\text{Skew} = \big|\overline{Q}{0} - \overline{Q}{1}\big|,
\end{equation}
where $\overline{Q}_g$ is the average total queue length (running + waiting) on GPU $g$. A smaller value indicates more balanced utilization, meaning neither GPU is persistently overloaded or underutilized.

We further decompose this metric into \emph{running skew} and \emph{waiting skew}. Running skew reflects imbalance in the number of active requests being processed on each GPU, capturing differences in effective throughput across models. Waiting skew captures imbalance in queued (not yet processed) requests, revealing cases where one GPU becomes a bottleneck and accumulates backlog.

\subsection{Arrival Patterns}
Our load generator supports three arrival patterns designed to mimic a broad spectrum of real-world traffic conditions: (1) Poisson, which approximates natural user arrivals; (2) micro-burst, which creates short, high-intensity spikes; and (3) sustained-overload, which maintains a consistently high load beyond the system’s nominal capacity. During cost-model training, we cycle through all three patterns to expose the serving system to diverse hardware states—ranging from completely idle GPUs to moderate load, burst-induced contention, and deep queue buildup. This diversity in system load and resource pressure helps the latency predictor learn how TTFT and TPOT vary under different hardware conditions, improving its robustness and generalization across serving scenarios.

For router evaluation, we use sustained-overload with swept arrival rates, as high-load regimes better reveal routing differences by forcing routers to balance queue pressure, model efficiency, and GPU heterogeneity.

\begin{figure*}[t]
    \centering
    \vspace{-0.1in}
    \includegraphics[width=\linewidth]{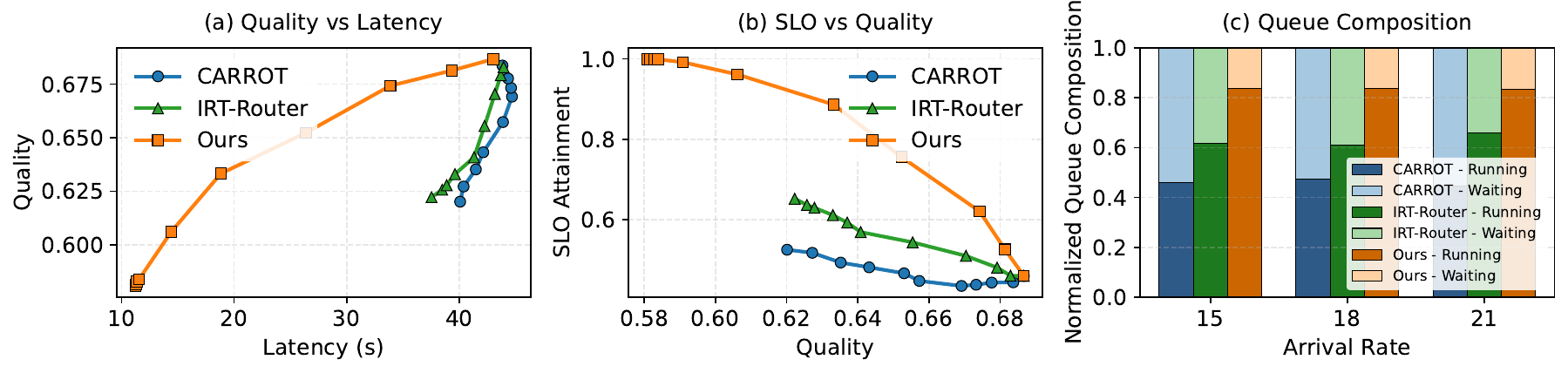}
    \caption{Performance comparison between HW-Router and CARROT across three key metrics. (a) Quality-latency Pareto frontier obtained by varying the trade-off parameter $\lambda$. (b) SLO attainment as a function of quality requirements. (c) Queue composition across different arrival rates, where HW-Router exhibits balanced load distribution with minimal waiting queues while CARROT suffers from excessive queueing, indicating poor resource utilization.}
    \label{fig:results}
\end{figure*}

\section{Evaluation}
We use the following three research questions (RQs) to evaluate our HW-Router.

\noindent\textbf{RQ1:} Can HW-Router achieve better quality-latency trade-offs compared to existing routers under different arrival rates?
% We examine whether incorporating hardware awareness enables more efficient routing decisions that improve response quality at comparable latencies or reduce latency while maintaining quality.

\noindent\textbf{RQ2:} Can HW-Router maintain consistent SLO compliance across different response quality requirements?

\noindent\textbf{RQ3:} Can HW-Router routing lead to better load balancing across the model pool?

\subsection{RQ1: Better Latency--Quality Tradeoff}

Figure~\ref{fig:results}~(a) demonstrates that HW-Router achieves better quality-latency trade-offs compared to baselines. The Pareto frontiers are obtained by adjusting the trade-off parameter $\lambda$ in the scoring function $(1-\lambda)\cdot Q - \lambda \cdot C$. While CARROT and IRT-Router exhibit nearly identical performance—both suffering from static cost assumptions—HW-Router shows clear superiority across the entire spectrum of operating points. At comparable quality levels, HW-Router consistently delivers lower latency: to achieve a quality score of 0.62, both CARROT and IRT-Router require approximately 40 seconds while HW-Router needs only 16 seconds, representing a 60\% latency reduction. The convergence of CARROT and IRT-Router curves highlights a fundamental limitation: despite IRT-Router's advanced quality modeling, its reliance on fixed cost estimates prevents it from exploiting runtime optimization opportunities. In contrast, HW-Router's improvement stems from its ability to accurately predict actual serving latency based on real-time system conditions, enabling it to identify truly efficient model-GPU combinations. The smooth progression of HW-Router's curve also indicates more predictable performance across different $\lambda$ values, allowing system operators to confidently select operating points that balance their specific quality and latency requirements.

\subsection{RQ2: Robust SLO Guarantees}

Figure~\ref{fig:results}~(b) reveals that HW-Router achieves higher SLO attainment rates compared to both CARROT and IRT-Router across varying quality requirements. While CARROT's SLO attainment hovers around 50\% regardless of quality target, indicating that half of all requests violate latency constraints and IRT-Router performs only marginally better, achieving approximately 65\% at lower quality levels. In contrast, HW-Router demonstrates exceptional SLO compliance, particularly at lower quality thresholds. At a quality target of 0.60, HW-Router achieves near-perfect SLO attainment (approximately 97\%), while CARROT manages only 51\% and IRT-Router achieves 65\%. The gap widens at moderate quality levels: at quality 0.63, HW-Router maintains 90\% SLO attainment compared to IRT-Router's 60\% and CARROT's 48\%. Even as quality requirements increase to 0.65, HW-Router still achieves 76\% SLO attainment, while both baselines fall below 50\%. Our higher SLO performance demonstrates that hardware-aware routing is essential for predictable latency guarantees—by accurately predicting actual serving times including queueing delays and resource contention, HW-Router can make informed decisions that avoid SLO violations. In contrast, static cost models lead to chronic underestimation of actual latency, resulting in consistent SLO violations that would be unacceptable in production deployments.

\subsection{RQ3: Balanced Load Distribution}

Efficient GPU scheduling requires not only selecting fast models, but also distributing work evenly across available GPUs. We evaluate both aspects using two complementary metrics, i.e., Queue Composition in Figure~\ref{fig:results}~(c) and GPU Load Skew in Table~\ref{tab:gpu-load-skew}.

% \noindent\textbf{Queue Composition.} Figure~\ref{fig:results}~(c) reveals the distribution of running versus waiting requests in each router's queue. CARROT exhibits a concerning pattern with waiting requests comprising 53-55\% of its total queue across all arrival rates, indicating that over half of the requests are stalled rather than being processed. This high waiting fraction indicates severe bottlenecks and poor model selection—requests accumulate in queues of overloaded models while other LLMs and GPUs remain underutilized. In stark contrast, HW-Router maintains approximately 83\% running requests with only 16\% waiting, demonstrating efficient resource utilization where GPUs actively process requests rather than accumulating backlogs. This 3$\times$ reduction in waiting fraction directly translates to lower queueing delays: requests in HW-Router's system spend most of their time in actual computation rather than waiting for resources. The consistency of these ratios across different arrival rates (e.g., 15, 18, 21) further validates that HW-Router's hardware-aware routing effectively prevents queue buildup even as load increases, whereas CARROT's static approach leads to persistent congestion regardless of arrival rate.

\noindent\textbf{Queue Composition.}
Figure~\ref{fig:results}~(c) shows the distribution of running versus waiting
requests for each router. CARROT consistently exhibits a problematic queue
profile: across arrival rates 15, 18, and 21, \textbf{53--55\%} of its total queue
is waiting, meaning more than half the requests are stalled rather than being
processed. IRT performs moderately better but still inefficiently, with
\textbf{34--39\%} of its queue in the waiting state, indicating that many requests
are still funneled into congested GPUs while others remain underutilized. In
sharp contrast, HW-Router keeps \textbf{83--84\%} of requests in the running state
and only \textbf{16--17\%} waiting, demonstrating substantially improved resource
utilization. This \textbf{3.1--3.4$\times$} reduction in waiting fraction directly reduces
queueing delays: requests spend the majority of time computing rather than
waiting for GPU availability. The stability of these ratios across all tested
arrival rates confirms that hardware-aware routing consistently prevents queue
buildup under increasing load, whereas both CARROT's static routing and IRT's
heuristic-based approach continue to suffer from persistent congestion.

\begin{table}[ht!]
\centering
\small
\vspace{-0.1in}
\caption{GPU Load-Balance Metrics Across Arrival Rates. 
Lower skew indicates better balancing.}
\label{tab:gpu-load-skew}
\begin{tabular}{cccccc}
\toprule
\textbf{Arrival} & \multirow{2}{*}{\textbf{Router}} & \textbf{Average} &
\textbf{Load} & \textbf{Running} & \textbf{Waiting} \\
\textbf{Rate $\rho$} &  & \textbf{Load} & \textbf{Skew} & \textbf{Skew} & \textbf{Skew} \\
\midrule
15 & CARROT      & 6.451 & 12.903 & 5.946 & 6.957 \\
15 & IRT-Router  & 6.466 & 12.930 & 7.989 & 4.941 \\
15 & HW-Router   & 7.001 &  2.089 & 4.999 & 2.911 \\
\midrule
18 & CARROT      & 6.458 & 12.916 & 6.103 & 6.814 \\
18 & IRT-Router  & 6.424 & 12.841 & 7.833 & 5.008 \\
18 & HW-Router   & 7.149 &  1.995 & 4.995 & 3.000 \\
\midrule
21 & CARROT      & 6.462 & 12.924 & 5.757 & 7.167 \\
21 & IRT-Router  & 6.321 & 12.640 & 8.313 & 4.327 \\
21 & HW-Router   & 7.154 &  1.691 & 4.740 & 3.049 \\
\bottomrule
\end{tabular}
\vspace{-0.1in}
\end{table}

\noindent\textbf{GPU Load Skew.}
Table~\ref{tab:gpu-load-skew} reports GPU load-balance metrics across all
arrival rates. CARROT continues to show extreme imbalance, with load skew
values around 12.9—meaning one GPU processes roughly $13\times$ more requests
than the other. This imbalance appears consistently in both running queues
(5.8--6.1) and waiting queues (6.8--7.2), confirming that CARROT persistently
overloads a single GPU while leaving the other underutilized.

IRT-Router reduces skew slightly compared to CARROT but remains fundamentally
unbalanced. Its load skew stays high (12.6--12.9), with running-queue skew often
$\geq 7.8$ and waiting-queue skew $\geq 4.3$, indicating that IRT's
priority-based heuristic still funnels most requests to one GPU and does not
correct imbalance under load.

HW-Router is the only router that actually fixes the problem. It cuts load skew
down to 1.7--2.1 across all arrival rates—an \textbf{84--87\% reduction}
relative to CARROT and an equally large improvement over IRT. Running-queue
skew falls to 4.7--5.0 (approximately 18\% lower than CARROT/IRT), and
waiting-queue skew drops to 2.9--3.0 (approximately 55--60\% reduction). These
balanced metrics show that HW-Router not only distributes active work evenly
but also prevents queue buildup on any single GPU. The slightly higher average
load (7.0--7.2 vs. CARROT's 6.4--6.5 and IRT's 6.3--6.5) indicates better
resource utilization—both GPUs are kept busy instead of one overloaded and one
idle. This behavior holds from arrival rate 15 through 21, demonstrating that
hardware-aware routing eliminates GPU hotspots even as system pressure
increases.

\section{Conclusion}

This paper introduces HW-Router, the first hardware-aware routing framework for multi-LLM serving that moves beyond static cost modeling. By incorporating real-time system signals—queue states, memory usage, and GPU-specific metrics—HW-Router predicts actual serving latency rather than relying on static model attributes. Our evaluation shows that HW-Router achieves \textbf{3.4--3.9$\times$} lower end-to-end latency, \textbf{46--48} percentage-point higher SLO attainment, and \textbf{6--8$\times$} lower GPU load skew compared to both CARROT and IRT, along with a \textbf{3.1--3.4$\times$} reduction in waiting-queue fraction. These gains are achieved with only $\sim$200~$\mu$s of additional routing time over prior routers. Overall, the results demonstrate that effective LLM routing in heterogeneous deployments requires dynamic, hardware-aware decision-making rather than static quality--cost trade-offs.



%\bibliographystyle{IEEEtran}
\newpage
\bibliographystyle{ACM-Reference-Format}
\bibliography{bib}

\end{document}